# Seesaw of Saltwater and Inundation Drives Methane Emissions in Coastal Tidal Wetlands

Xun Cai[1], Xiucheng Yang[2], and Peter A. Raymond[1]

[1]Yale School of the Environment, Yale University, New Haven, CT, 06511, USA.

[2]Department of Natural Resources and the Environment, University of Connecticut, Storrs, CT, 06229, USA

Corresponding author: Xun Cai (xun.cai@yale.edu); ORCID: 0000-0002-4251-2384

Additional author information:

Xiucheng Yang (xiucheng.yang@uconn.edu); ORCID: 0000-0001-5134-9614

Peter A. Raymond (peter.raymond@yale.edu); ORCID: 0000-0002-8564-7860

**Abstract**

Wetlands are significant carbon sinks, yet methane emissions partially offset this function due to its high global warming potential. Coastal tidal wetlands, unlike non-tidal wetlands, are regulated by oceanic drivers like salinity gradients and tidal inundation, which strongly influence methane production and release but remain poorly represented in regional assessments. Here, we estimate methane emissions from U.S. East Coast tidal marshes, by integrating ocean model, remote sensing datasets, empirical relationships from metadata. Spatially, emissions reflect the combined effects of marsh extent and per-unit-area flux rates, with hotspots occurring under lower salinity, higher inundation, and lower latitudes. Temporally, temperature and salinity dominate decadal-scale interannual variability. Between 2001 to 2020, total methane emissions are estimated at 0.019–0.038 Tg yr$^{-1}$, with local fluxes rate ranging from 0 to 20 g m$^{-2}$ day$^{-1}$. Following pronounced hydrological variability in the early 2000s, emissions have increased steadily since 2007 at approximately 802 t yr$^{-1}$, driven by warming, freshening, and enhanced inundation. Projections under IPCC climate scenarios indicate that increasing inundation will amplify methane emissions with sea-level rise, until a threshold near 0.75 m SLR, beyond which saltwater intrusion increasingly suppresses further growth, highlighting the critical role of salinity–inundation interactions in coastal methane dynamics.

# 1 Introduction

Methane is a potent greenhouse gas with a global warming potential approximately 30 times greater than that of $CO_2$ over a 100-year timescale (Masson-Delmotte, 2021). Methane emissions from wetlands contribute significantly to atmospheric levels, with inland freshwater wetlands recognized as key methane sources in many studies (Bridgham et al., 2006; Zhang et al., 2017; Saunois et al., 2020; Rosentreter et al., 2021). Meanwhile, coastal tidal wetlands (e.g., tidal marshes and mangroves) also play an important role in regional and global methane budgets, and their emissions are uniquely influenced by oceanic processes such as tidal inundation and saltwater intrusion (Rosentreter et al., 2021). These intertidal ecosystems are characterized by highly dynamic hydrological regimes driven by tides, saltwater intrusion, and episodic events such as storm surges, all of which affect methane production and emission pathways (Poffenbarger et al., 2011; van Dijk et al., 2015; Minick et al., 2019).

In wetland soils, methane is primarily produced through anaerobic decomposition of organic matter through microbial processes under oxygen-limited conditions (Megonigal and Schlesinger, 2002; Kirschke et al., 2013). In tidal wetlands, factors such as soil temperature, water table fluctuations, salinity, and vegetation type strongly influence methane production and emission rates (Fig. 1; Mitsch et al., 2013; Poffenbarger et al., 2011; Cui et al., 2024). For example, elevated temperatures tend to enhance methane emissions by accelerating microbial metabolic rates, including methanogenesis (Vizza et al., 2017). Similarly, greater inundation promotes methane production by increasing the extent and duration of anoxic conditions in wetland sediments (Calabrese et al., 2021). Saltwater intrusion, however, also introduces sulfate, promoting sulfate reduction over methanogenesis and thus suppressing per-unit-area methane fluxes in tidal wetlands (Fig. 1; Chambers et al., 2011; van Dijk et al., 2015; Minick et al., 2019).

Recent research has highlighted that despite lower per-unit-area fluxes compared to inland freshwater wetlands, the extensive spatial coverage of coastal tidal wetlands still makes them significant contributors to regional greenhouse gas emissions (Rosentreter et al., 2021).

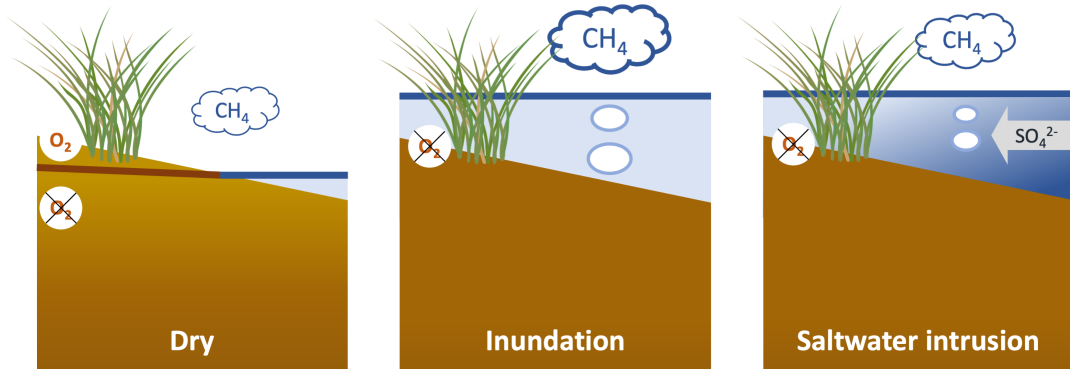

**Fig. 1.** Conceptual diagram of methane emission processes in marsh wetlands under exposed, inundated, and saltwater intrusion conditions.

While numerous studies have examined methane fluxes from coastal tidal wetlands at site and plot scales, often through chamber-based or eddy covariance measurements, regional-scale patterns and drivers remain less characterized (Al-Haj et al., 2020; Rosentreter et al., 2021; Arias-Ortiz et al., 2024). Most regional and global studies of wetland-driven methane emissions primarily focus on inland freshwater systems, often overlooking coastal tidal wetlands (Bansal et al., 2023; Chen et al., 2024; Deng et al., 2024; Xiao et al., 2024; Yuan et al., 2024; Bernard et al., 2025; Chen et al., 2025; He et al., 2025; Ury et al., 2025; Ying et al., 2025; Zhang et al., 2025; Zhu et al., 2025). This knowledge gap limits our ability to accurately estimate wetland methane emissions at larger spatial scales and under changing climate conditions. Here we present a regionally scaled, integrated assessment that combines field measurement metadata, remote sensing products, and ocean modeling to quantify methane emissions from U.S. East Coast tidal wetlands, focusing on tidal marshes. We examine decadal trends and regional drivers of methane emission variability, linking the spatiotemporal patterns to dominant processes. Furthermore, we project methane emission changes under warming and sea-level rise (SLR) scenarios, providing insights to support coastal management strategies.

## 2 Spatial patterns of methane emissions across coastal tidal marsh wetlands

The spatial distribution of methane emissions along the U.S. East Coast is shaped by the combined influence of two key factors: the extent of marsh area and the local per-unit-area methane flux rates (Fig. 2a). Emission hotspots emerge where extensive wetland coverage coincides with elevated fluxes, resulting in disproportionately high total emissions. Notable examples include the Salem River Wildlife Management Area in Delaware Bay (Fig. 2c) and the Blackwater National Wildlife Refuge in Chesapeake Bay (Fig. 2d), where this synergistic overlap is most pronounced. In contrast, areas with either limited marsh extent or moderate-low flux rates exhibit lower emissions, contributing to the spatial heterogeneity observed across the region.

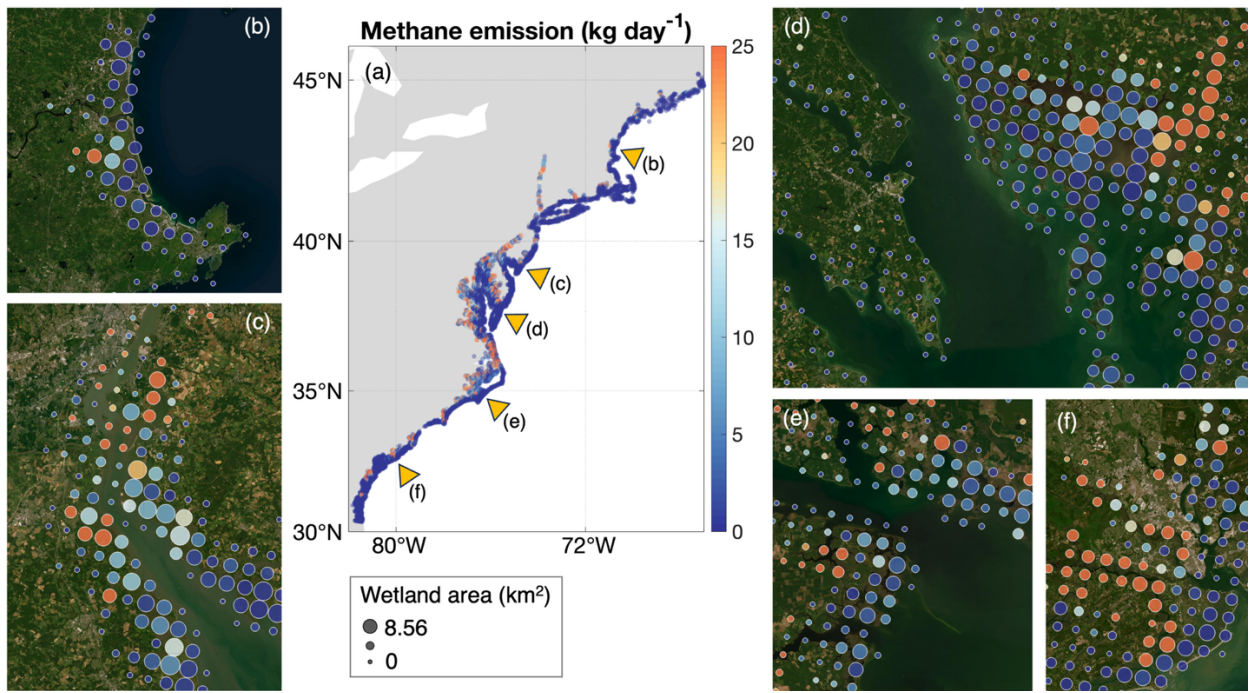

**Fig. 2.** (a) Spatial distribution of mean methane emissions per 9 km² sampling unit. (b–f) Zoom-in maps of (b) Plum Island Estuary, (c) Delaware Bay, (d) Chesapeake Bay, (e) Albemarle-Pamlico Sound Estuary, and (f) Cooper River Estuary. In panels (b–f), circle sizes reflect the wetland area per sampling unit based on remote sensing data. Colors in these panels represent total emissions per sampling unit.

In terms of marsh extent, the most tidal marsh area in the study region is concentrated in the South Atlantic Coast (55%) and Mid-Atlantic Coast (40%), whereas the New England Coast accounts for only a small fraction (5%) (Figs. 3b-d and S1b). The relatively small fraction at the New England Coast underscores the dominant role of the Mid- and South Atlantic regions in shaping total regional methane emissions. However, methane emissions are not proportionally distributed across these regions. The South Atlantic Coast, Mid-Atlantic Coast, and New England Coast contribute 60%, 37%, and 3.4% of the total methane fluxes, respectively (Fig. 3b-d). Thus, the South Atlantic Coast contributes to a disproportionately large share of total emissions relative to its areal extent. This discrepancy reflects spatial differences in per-unit-area flux rates, which are influenced by regional variations in environmental drivers

Temperature patterns contribute to these latitudinal gradients, with generally higher flux rates in southern estuaries that experience warmer climates and therefore higher microbial activity, and methane production (Fig. S1cd). Elevated temperatures support higher rates of primary production and anaerobic decomposition in marsh soils (Inglett et al., 2012; Fig. S1). Additionally, the longer growing seasons characteristic of southern regions may further amplify methane emissions, as suggested by prior studies across diverse ecosystems (e.g., Treat et al., 2018). However, the empirical framework used in this study does not explicitly incorporate growing season length or vegetation productivity dynamics, which may introduce some uncertainty, particularly a potential underestimation of methane fluxes in the southernmost marshes.

Elevated flux rates are also evident in regions characterized by prolonged inundation (e.g., regions of higher tidal range) coupled with reduced salinity exposure (e.g., landward regions), such as the upper reaches of Chesapeake Bay, and the landward zones of the Albemarle–Pamlico Estuarine System (APES) (Fig. S1cef). Prolonged inundation promotes anoxic soil conditions favorable for methanogenesis (Herbert et al., 2015), while reduced salinity limits competition from sulfate-reducing bacteria, further enhancing methane production (Selak et al., 2025). At the same time, methane produced in deeper sediments typically encounters an oxic surface layer where methanotrophs can consume a portion of the gas before it reaches the atmosphere (Roslev and King, 1996). When inundation increases and the oxic layer becomes thinner, methanotrophic consumption is reduced, making a larger fraction of methane emit (Cui et al., 2024). Taken together, these processes create more favorable conditions for methane production and less capacity for methane removal, resulting in amplified emissions in low-salinity, frequently inundated marsh zones. Conversely, areas exposed to higher salinities generally display lower flux rates, consistent with the suppressive effects of sulfate-rich conditions on methane production (Fig. S1ce).

Together, these spatial patterns highlight how both the distribution of wetland area and environmental controls on flux intensity shape the overall geography of coastal methane emissions (Fig. 2a). Quantitatively, estimated regional methane emissions ranged from approximately 0.019 to 0.038 Tg yr$^{-1}$ (Figs. 2 and 3b). Specifically, the total coastal tidal marsh wetland area represented in this study is about 7,200 km², which is roughly 13% of the global coastal tidal salt marsh wetland extent (Campbell et al., 2022). Estimated flux rates range from 0 to 20 g m$^{-2}$ day$^{-1}$ (≈ 0.45 Tmol CH$_4$–C yr$^{-1}$; Fig. S1c), which is comparable to estimates by Al-Haj and Fulweiler (2019). Based on global estimates from Rosentreter et al. (2021), coastal tidal marshes emit a median of 0.18 Tg yr$^{-1}$ of methane (range 0.020–0.89 Tg yr$^{-1}$), so the 13 percent share in our study region corresponds to roughly 0.023 Tg yr$^{-1}$ (0.0026–0.12 Tg yr$^{-1}$), which aligns well with our regional estimates. However, the global mean estimates reported by Rosentreter et al. (2.00 ± 1.51 Tg yr$^{-1}$) is much higher than the median, and a proportional 13 percent share of that value would exceed the emissions estimated for our study region.

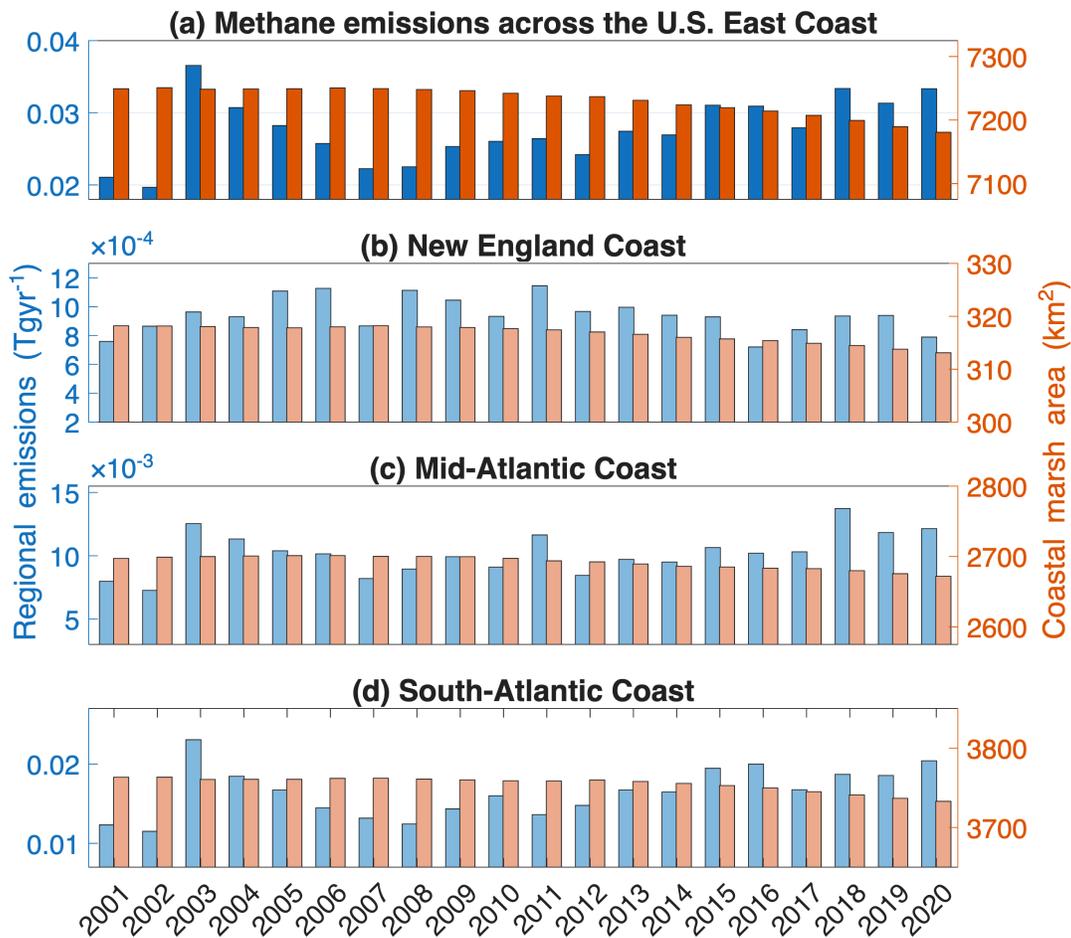

**Fig. 3.** Annual total methane emissions from 2001 to 2020 (a) across the U.S. East Coast marshes, which is composed of (b) New England Coast, (c) Mid-Atlantic Coast, and (d) South-Atlantic Coast.

## 3 Temporal variability of methane emissions across regions

On the interannual scale, methane emissions from U.S. East Coast marshes show distinct interannual variability for the whole region and major sub-regions (Fig. 3b-d). Over the two-decade period, interannual variability in methane emissions strongly corresponds to fluctuations in temperature (Fig. 4abc-1) and salinity (Fig. 4abc-3) compared to inundation (Fig. 4abc-2), while inundation exhibits comparatively lower variability. This limited variability in inundation likely reflects the slow progression of SLR and gradual morphological adjustments of marshes, which induce only minor year-to-year changes in flooding conditions. Principal Component Analysis (PCA) further supports this interpretation, highlighting temperature and salinity as the dominant drivers of methane flux interannual variability. Inundation contributes minimally to the first two principal components due to its relative temporal stability (Figs. S2–S3). Regression of methane emissions against PCA scores reveals that the first principal component (PC1), which captures strong contributions from temperature and moderate input from salinity, explains over 50% of the variance across all subregions ($R^2 > 0.50$). Complementary Random Forest out-of-bag (OOB) variable importance analysis reinforces this pattern, ranking temperature (0.48–0.57) and salinity (0.39–0.48) as the most influential predictors of methane fluxes, whereas inundation consistently ranks lowest (0.03–0.04) in predictive importance (Fig. S4). Interannual variability in salinity, largely controlled by freshwater discharge under varying hydrological conditions, therefore plays a key role in regulating methane emissions (Cai et al., 2025). Together, these statistical analyses confirm that regional-scale decadal interannual variability in methane emissions is primarily governed by changes in temperature and salinity, with inundation exerting a secondary influence during this timeframe.

The interannual variability and dominant drivers of methane emissions also reflect substantial spatial heterogeneity across coastal subregions. While temperature consistently emerges as the leading driver of interannual variability across all regions, Random Forest out-of-bag (OOB) importance analysis reveals regional differences in secondary controls. For example, salinity exerts a notably stronger influence in the Mid-Atlantic Coast (importance score: 0.48) compared to other regions (e.g., 0.41), and in this subregion, salinity is nearly as influential as temperature (Fig. S4). This heightened influence of salinity in the Mid-Atlantic is likely driven by its unique hydrogeomorphic context: large estuaries such as Chesapeake Bay and Delaware Bay receive substantial freshwater inflows and are characterized by broad transition zones where dynamic saltwater intrusion interacts with extensive estuarine wetlands. This contrasts with the smaller, lagoonal estuaries of the South and New England coasts, which exhibit less pronounced salinity gradients and estuarine circulation.

Further evidence of spatial heterogeneity is revealed through PCA, particularly the third principal component (PC3), which explains more variance in methane emissions than the second principal component (PC2) despite accounting for a smaller portion of total driver variability (12.67%). In the New England Coast, PC3 is primarily shaped by salinity, whereas in the Mid-Atlantic and South Atlantic Coasts, inundation is the dominant component. PC3 explains 22.36%, 13.21%, and 7.63% of methane emission variability in the Mid-Atlantic, South Atlantic, and New England regions, respectively (Figs. S5–S6). These findings imply that methane emission variability in the New England Coast is more sensitive to changes in salinity, while in the southern regions, it is more responsive to fluctuations in inundation conditions. This may reflect the greater proportion of high marshes in New England region, where less frequent inundation happens (Hopkinson et al., 2025). Overall, these regional differences underscore the importance of accounting for spatial heterogeneity in environmental drivers when assessing methane emissions. They also highlight that decadal-scale emission patterns across U.S. East Coast tidal marshes are shaped by the interplay of locally dominant processes, ranging from temperature and salinity dynamics to estuarine morphology and hydrological gradients.

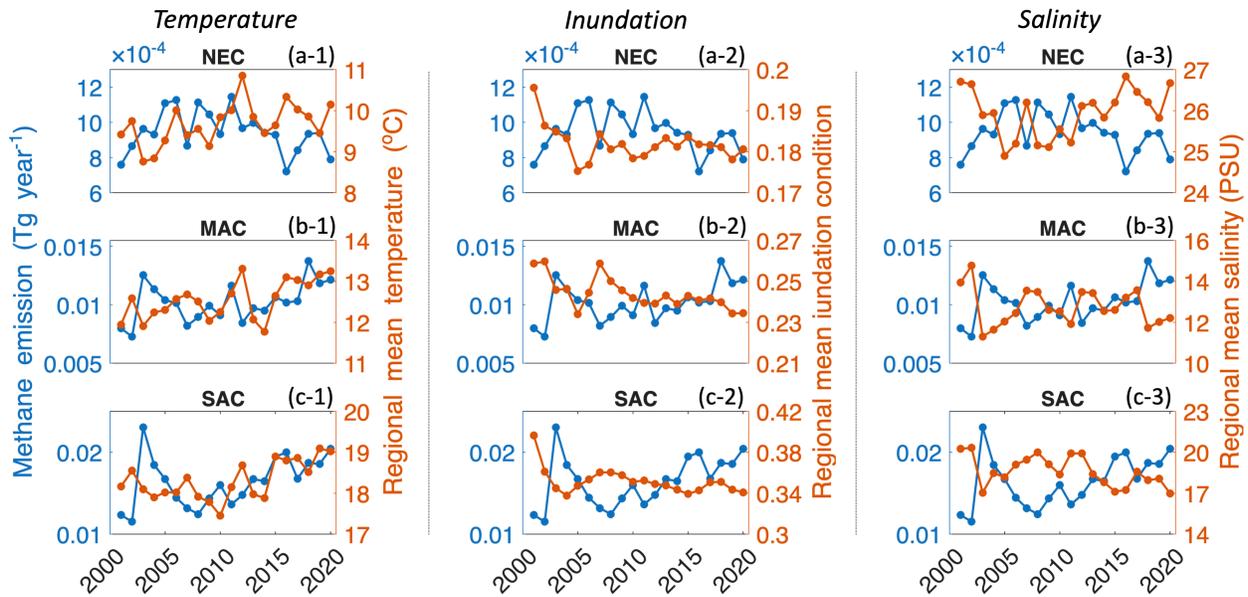

**Fig. 4.** Annual mean methane emissions in relation to (1) temperature, (2) inundation conditions, and (3) salinity from 2001 to 2020 across U.S. East Coast marshes, shown for the (a) New England Coast (NEC), (b) Mid-Atlantic Coast (MAC), and (c) South Atlantic Coast (SAC).

Besides interannual variability, methane emissions and flux rates exhibit a clear seasonal cycle, with low rates in winter and pronounced increases beginning in the spring, peaking during June–August, and declining again in fall (Fig. 5). This seasonal pattern reflects the dominant influence of temperature, which accelerates plant and microbial activity during warmer months, resulting in both higher median fluxes and greater variability. The wider range of emissions, especially during summer, also indicates the influence of secondary drivers such as salinity exposure and hydrological variability. During low-discharge summer months, intensified saltwater intrusion may suppress methane emissions, whereas storm-driven freshening and increased inundation can enhance methane release (Poffenbarger et al., 2011; Arias-Ortiz et al., 2024). For example, in July, median flux rates increase from 4 to 22 and 60 g m$^{-2}$ day$^{-1}$ when transitioning from high to moderate and low salinity conditions. Likewise, median emissions rise from 0.01 to 1 and 75 g m$^{-2}$ day$^{-1}$ under low, moderate, and high inundation conditions, respectively (Fig. 5bc). These large contrasts in median values and spread across environmental gradients contribute substantially to seasonal variability. Furthermore, in upland areas with low inundation that are rarely flooded and more strongly linked to wet and dry hydrological conditions, episodic inundation events can trigger methane emissions (Fig. 5c-1). These emissions occur most frequently in September rather than during the summer months, reflecting the influence of preceding summer drought conditions (Fig. 5c-1). These seasonal dynamics at different salinity and inundation zones indicate the temperature-dominant pattern is modulated by the competition between saltwater intrusion and inundations conditions, in shaping seasonal methane fluxes.

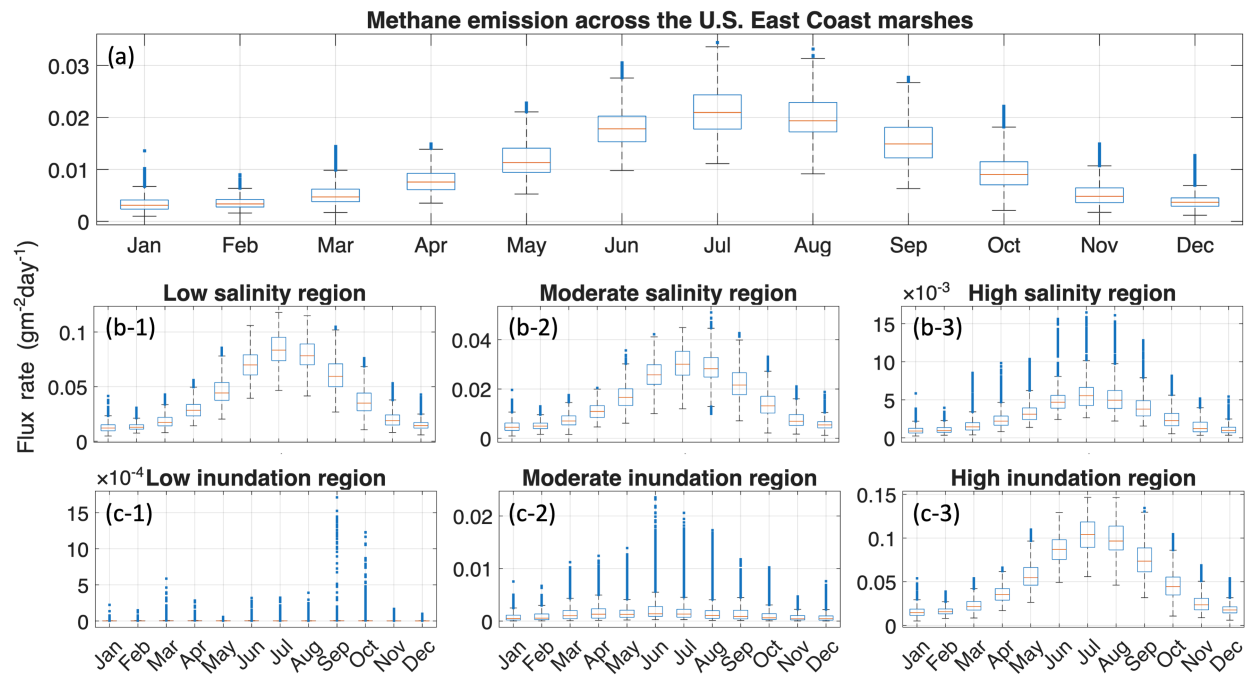

**Fig. 5.** Boxplots of seasonal mean methane flux rates (g m$^{-2}$ day$^{-1}$) (a) across U.S. East Coast marshes and stratified by local mean (b) salinity and (c) inundation conditions. In panels (b) and (c), low, moderate, and high categories correspond to the lower, middle, and upper 33.33 percent of values across the entire region, respectively.

**4 Total fluxes and decadal trends in methane emissions from U.S. East Coast marshes**

Methane emissions from U.S. East Coast marshes show no consistent long-term increasing or decreasing trend between 2001 and 2020 (Fig. 3a). The early 2000s (2001–2006) were characterized by stable marsh extent and pronounced hydrological fluctuations, including 2002 (the driest year since the 1930s) and 2003 (the second wettest year since 1930s), followed by several years of wet-to-neutral conditions (Cai et al., 2025). These contrasting climate and hydrological regimes led to significant divergences in salinity patterns (Fig. 4abc-1; Cai et al., 2025) and altered inundation conditions in upstream wetlands (Fig. 4abc-2), both of which are key controls on methane fluxes (Poffenbarger et al., 2011; Arias-Ortiz et al., 2024). From 2007 to 2020, methane emissions from U.S. East Coast tidal marshes exhibited a modest but consistent upward trend, increasing by approximately 803 t yr$^{-1}$ (Fig. 3a), accompanied by reduced interannual variability compared to the early 2000s. During this period, marsh extent, which had remained relatively stable in the early 2000s, began to decline gradually at a statistically significant rate of 5 km² yr$^{-1}$ across all three subregions (Fig. 3a, Fig. S5). Despite this areal contraction, total regional emissions continued to rise, indicating that increasing per-unit-area methane fluxes have outpaced the loss in the wetland area. This trend is particularly pronounced in the marsh-dense Mid-Atlantic and South Atlantic Coasts, which showed significant emission increases of 296 and 556 t yr$^{-1}$, respectively (Fig. 3b-d). This decoupling of wetland extent and emission trends underscores that changes in environmental drivers, such as temperature, salinity, and inundation, have elevated per-area methane fluxes, making them the dominant factor behind recent regional-scale emission increases.

The sustained upward trend in methane emissions over the past 15 years coincides with region-wide environmental changes, rising temperatures, increased inundation, and decreasing salinity, particularly in the Mid-Atlantic and South Atlantic Coasts (Fig. 4). Temperature trends over this period show regional increases of 0.04 °C yr$^{-1}$ in the New England Coast (p = 0.15), 0.06 °C yr$^{-1}$ in the Mid-Atlantic Coast (p = 0.063), and 0.08 °C yr$^{-1}$ in the South Atlantic Coast (p = 0.012). Although not all trends are statistically significant, the increasing temperatures support higher methane production. Inundation has also intensified, as indicated by negative trends in the inundation index (where more negative values represent greater inundation): -9 × 10$^{-5}$ yr$^{-1}$ (p = 0.38) for New England, -0.0010 yr$^{-1}$ (p = 0.0031) for Mid-Atlantic, and -0.002 yr$^{-1}$ (p = 0.0031) for South Atlantic regions. These changes suggest increases in tidal water inundation over time, which is especially true in the southern regions. The South Atlantic Coast exhibits a significant salinity decline of -0.2 PSU yr$^{-1}$ (p = 0.011), while the Mid-Atlantic shows a weaker, non-significant decrease of -0.06 PSU yr$^{-1}$ (p = 0.19). In contrast, the New England Coast displays a modest but significant increase of 0.08 PSU yr$^{-1}$ (p = 0.021), likely tied to reduced freshwater inputs. Although regional patterns differ, the broader shift toward lower salinity in marsh-dense areas intensifies methane production and contributes to higher emissions. Collectively, these trends of warmer temperatures, enhanced inundation, and lower salinity in key regions create favorable conditions for increased methane emissions, reinforcing the observed regional-scale upward trend since 2007.

## 5 Impacts of projected climate change on coastal tidal marsh wetland methane emissions

Building on our regional scale estimates of methane emissions from coastal tidal marshes, we assessed potential future changes by incorporating projected changes in temperature, salinity, and inundation due to SLR. Using outputs from an ocean model for salinity and inundation under SLR, alongside projected temperature increases, we evaluated how each driver, individually and in combination, could influence methane emissions by 2100 (Fig. 6a). These projections were applied within the framework of our simplified empirical model, assuming no change in the wetland area and excluding processes not explicitly represented (e.g., marsh migration, soil moisture dynamics). Under these assumptions, total methane emissions are estimated to rise to 80–110 T day$^{-1}$ by the end of the century across different climate scenarios, driven by the combined effects of warming, saltwater intrusion, and enhanced inundation (Fig. S6). While these estimates provide directional insight, precise quantification remains uncertain and will require more comprehensive, process-based modeling in future research.

From our analysis, both higher temperatures and increased inundation promote methane emissions by stimulating microbial activity and extending anaerobic conditions, while saltwater intrusion exerts an opposing, suppressive influence. Among these factors, temperature remains the dominant driver, amplifying methane fluxes and explaining the persistent upward trajectory of emissions, which is consistent with the multidecadal driver analysis presented above. Over century-scale projections, the influence of inundation on methane emissions has become substantially more pronounced compared to the past two decades. As sea level rises, inundation conditions intensify, resulting in longer and more frequent marsh submergence (Fig. S6a). These scenarios assume that U.S. coastal tidal marshes keep pace with rising seas, such that the increase in inundation reflects a larger tidal range, consistent with findings from previous studies on SLR impacts (Cai et al., 2022; Cai et al., 2023). This expanded hydroperiod creates more anaerobic environments favorable for methanogenesis, thereby amplifying methane emissions, especially under moderate SLR of ~0.5–1.0 m (Fig. 6a).

Focusing on the suppressive role of saltwater intrusion, century-scale estimations highlight the dynamic competition between inundation-driven enhancement and salinity-driven suppression of methane fluxes (Figs. 4b, S6a). Our analysis indicates that a SLR of approximately 0.75 m represents a critical threshold, at which salinity increases by about 1.50 PSU and becomes a dominant factor in counteracting tidal inundation (Fig. 6b). Beyond this point, the influence of saltwater intrusion increasingly outweighs inundation enhancement, suppressing methanogenesis, and gradually slowing the acceleration of methane emissions. However, even when saltwater intrusion dampens methane production at higher SLR, warming sustains elevated flux rates, as shown by sensitivity analyses under temperature increments of +1.50°C, +3.00°C, and +5.30°C (Fig. S6b). This suggests that temperature operates as a persistent multiplier of flux intensity and highlights the nonlinear response of methane fluxes to Earth system projections.

In conclusion, the century-scale estimations reveal a shift in the balance of dominant drivers: inundation, relatively stable and minor over recent decades, emerges as a more important regulator under rising seas, while saltwater intrusion increasingly constrains methane release at higher SLR thresholds. Yet temperature continues to exert the strongest and most consistent influence. These findings indicate that future methane fluxes from coastal tidal marshes will be shaped by the interplay of hydrological and thermal drivers, with regional variability and thresholds determining whether inundation or salinity exerts greater control.

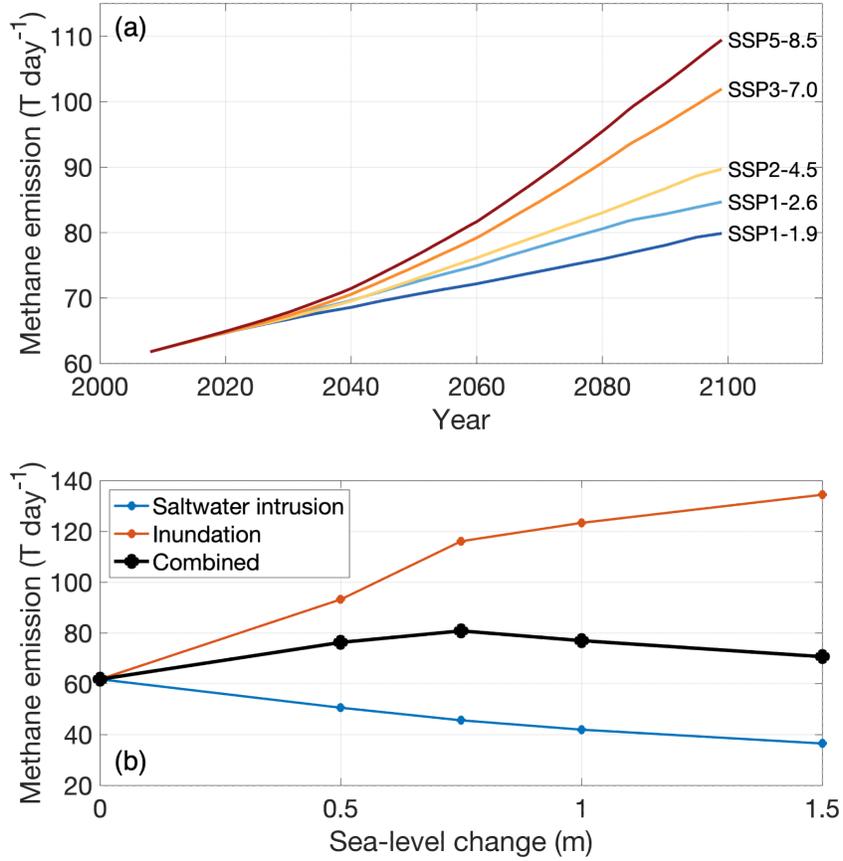

**Fig. 6.** (a) Predicted methane emissions at the U.S. East Coast marshes under five scenarios of warming and SLR based on IPCC AR6. (b) Predicted methane emissions at the U.S. East Coast marshes under SLR with competition between drivers of enhanced saltwater intrusion and increasing inundation associated with larger tidal range.

**Methods**

**1 Available data**

The development of empirical relationship is based on four independent eddy covariance (EC) tidal marsh datasets at U.S. East Coast available through AmeriFlux (US-StJ, US-MRM, US-RRC, and US-NC4We also utilized the metadata generated by Arias-Ortiz et al. (2024) to further support this study, specifically for the empirical relationship between salinity and methane fluxes. This dataset includes 35 contributed $CH_4$ flux sub-datasets of 122 tidal marsh sites measured with chambers across CONUS (Arias-Ortizet al., 2024). Each dataset follows established standards and quality assurance/quality control procedures detailed in previous studies (Chu et al., 2023; Arias-Ortizet al., 2024). Overall, there are 510 chamber date points containing measurements with $CH_4$ flux rate along with temperature, daily inundation conditions, and salinity. For each single predictor of temperature, inundation, and salinity, there are in total 7553, 1735, and 1030 data points respectively that are generally evenly distributed to support the estimation of empirical relationship.

**2 Ocean model and scenarios at the North American Atlantic Coast**

We employed the three-dimensional unstructured-grid NAAC (v1.0) model to provide the necessary salinity and inundation data for this study (Cai et al., 2025). The model spans the Gulf of Maine, Mid-Atlantic Bight, and most of the South Atlantic Bight along the North American Atlantic Coast. Built upon the Semi-implicit Cross-scale Hydroscience Integrated System Model (SCHISM; Zhang et al., 2016), NAAC (v1.0) offers a validated 20-year simulation (2000–2020) with coastal grid resolutions of finer than 200 meters. Its seamless connection from the coastal ocean to tidal marshes enables detailed representation of hydrodynamic processes, including saltwater intrusion and tidal inundation, supporting regionally consistent methane emission estimates in this study.

## 3 Coastal tidal marsh wetland annual cover map

We utilized U.S. coastal tidal marsh wetland cover maps derived from Landsat satellite observations (Yang et al., 2022). This dataset provides spatially and temporally consistent maps of tidal wetland dynamics at high resolution (30 m) spanning nearly four decades. It was developed using dense time-series analysis combined with a machine learning–based classification approach and has been validated through cross-comparisons with other regional databases (e.g., Campbell et al., 2020), ensuring reliability across sub-regions within the domain. We constrain our study area to regions north of 30.5° N to focus on coastal tidal marsh while excluding mangrove-dominated coastal landscapes. In our analysis, we applied the mapped tidal wetland locations to extract corresponding hydrological variables from the ocean model and used the wetland coverage as the spatial basis for aggregating methane emission estimates. This method allowed us to accurately scale emissions to the actual wetland extent within each grid cell. Incorporating this satellite-based dataset provided consistent, up-to-date wetland area information across the entire study region over the past two decades, which was essential for producing robust and regionally representative methane flux estimates.

## 4 Process-Based Bottom-Up Model for Coastal tidal marsh wetland Methane Emissions

We build the methane emission model at each grid as:

$$F_{\mathrm{CH_4}} = M_{\mathrm{CH_4}} \cdot A_w \cdot R(T, S, I) \tag{1}$$

where $F_{CH_4}$ is methane flux (g CH4 day$^{-1}$) and $M_{CH_4}$ is the molar mass of methane (g CH4 mol$^{-1}$). $A_w$ is wetland area of a pixel (m²). $R(T, S, I)$ is methane flux rate (mol C m$^{-2}$ day$^{-1}$) estimated by temperature $T$ (°C), salinity $S$ (PSU), and inundation patterns $I$, and is described as:

$$R(T, S, I) = R_0 \cdot C(T) \cdot F(S) \cdot M(I) \tag{2}$$

where $R_0$ (mol C m$^{-2}$ day$^{-1}$) is a reference rate and $C(T)$ is partially a Gaussian-type probability distribution describing the influence of temperature on methane fluxes, representing the combined effects of local production and methanogenesis processes governed by a constant rate parameter $k$:

$$C(T) = e^{k \cdot T} \tag{3}$$

and $F(S)$ describes the regulations from saltwater intrusion, where parameters constant $a$ is a negative value that is calibrated with observations:

$$F(S) = e^{a \cdot S} \tag{4}$$

The inundation pattern is determined by the daily inundation ratio value where $I$ whose range is from 0 (inundated) to 1 (exposed). Conditions of exposure leads to oxygen-rich layer that is associated with methane oxidation and decrease of methane emissions while conditions of inundation tends to prompt methane releases. $M(I)$ is expressed as:

$$M(I) = e^{b \cdot (1 - I)} \tag{5}$$

where parameter $b$ amplifies this coefficient when the submergence is prolonged.

The values of parameters $R_0$, $k$, $a$, and $b$ are estimated by the available eddy covariance monitoring and metadata collection (Arias-Ortiz et al., 2023; Fig. S8). The estimated methane emissions are validated with available monitoring data from four coastal eddy covariance stations (Fig. S9).

Methane flux parameterization was derived by fitting exponential functions to observational datasets from four eddy covariance towers and complementary chamber measurements along the U.S. East Coast (Fig. S8). These empirical relationships were then used to parameterize methane emissions across the broader regional model. Importantly, the parameter sets were calibrated using the combined observational record rather than site-specific tuning, ensuring general applicability across diverse marsh systems but potentially smoothing over regional differences

To support regional-scale estimates of methane emissions, we integrated multiple complementary components into a unified framework. Methane flux observations from eddy covariance towers and chamber measurements were first used to derive empirical relationships linking fluxes to key environmental drivers (temperature, inundation, and salinity), which were then validated against independent observational records. These driver fields were generated at high spatial and temporal resolution using a coastal hydrodynamic model (Cai et al., 2025), while annual tidal marsh areas were quantified from remote sensing analyses (Yang et al., 2022). By combining modeled drivers with remote sensing of marsh extent, the empirical functions were scaled up to produce regional methane emission estimates across the U.S. East Coast. This framework leverages the strengths of observational data, process-based modeling, and satellite products, allowing for a consistent and data-informed approach to quantify methane emissions at scales not achievable through site-level observations alone.

The model was validated against a time series of methane fluxes from the four eddy covariance sites (StJ, RRC, NC4, Mrm), representing distinct estuarine settings (Fig. S9).

Overall, the model reproduced the observed seasonal and interannual variability, particularly capturing the magnitude of peak emissions at sites such as NC4 and RRC. However, notable discrepancies remain. At StJ, simulated fluxes are underestimated, likely due to the ocean model resolution not resolving the lower bay branch where the tower is located, leading to an overestimation of local salinity. Nonetheless, StJ also exhibits a smaller observed flux range compared to the other sites, reducing its influence on regional scaling. At NC4, the model did not simulate methane uptake processes, which explains part of the observed mismatch, though it successfully reproduced the upper emission envelope. At the NC4 site, approximately 17.90% of the available data indicate methane uptake, which offsets about 10.10% of the total emissions. In contrast, the other eddy covariance stations do not show uptake of comparable magnitude. Although our current estimation framework does not explicitly incorporate methane uptake, the resulting uncertainty introduced by this omission is relatively small at the regional scale. Taken together, these results suggest that while the framework captures the broad magnitude and dynamics of methane emissions, uncertainties remain at finer spatial scales and for processes such as uptake, which warrant further refinement in future modeling efforts.

## 5 Implementation of Climate Scenarios

To evaluate the effects of warming and SLR on coastal methane emissions, we first conducted six one-year numerical experiments using SLR levels of 0, 0.5, 0.75, 1.0, and 1.5 m relative to the baseline ocean model. These SLR values represent a range of future projections consistent with greenhouse gas emission scenarios outlined in IPCC AR6 (Allan et al., 2023). The year 2008 was selected as the baseline because it reflects near-average hydrological conditions, neither particularly wet nor dry. Each simulation included a one-year spin-up period to minimize the influence of initial conditions.

We then incorporated warming scenarios of +0.2°C to +5.3°C (0.2, 0.5, 1.0, 1.5, 2.0, 2.5, 3.0, 3.5, 4.6, and 5.3°C), combined with corresponding salinity and inundation results from the SLR scenarios from the ocean model, to assess methane emission responses under various SLR–temperature combinations. In total, 50 coupled SLR–warming scenarios were simulated. Finally, methane emission projections for each future year were derived by integrating the IPCC AR6 scenario probabilities with model-based methane flux predictions across the 60 experiments.

## 6 Uncertainties and future improvements

A key limitation arises from the assumption of uniform empirical parameter sets across all regions. In reality, marsh systems differ substantially in vegetation species, sediment composition, and hydrological connectivity, all of which may influence the magnitude and temperature sensitivity of methane fluxes. By applying identical relationships across systems, we may miss critical regional distinctions in emission responses. Site-specific calibration or the use of regionally stratified parameterizations, potentially informed by expanded flux tower and chamber observations, would improve the accuracy of future estimates and allow for more nuanced projections.

Another uncertainty in our analysis is the relative dominance of temperature compared to other drivers such as salinity and inundation. While our results consistently show temperature as the strongest predictor of methane emissions, this may partly reflect the sensitivity of empirical models to thermal forcing and the relatively limited variability of inundation over the past two decades. It is possible that we overstate temperature's role in some contexts, particularly given the strong nonlinearity of salinity and inundation effects under future SLR. Future work with mechanistic modeling frameworks might explicitly couple thermal sensitivity with hydrological and biogeochemical processes to better capture the interactions among drivers and between systems.

Our analysis also omits methane uptake, which can reduce net emissions through microbial oxidation. While we expect this process to be less influential at seasonal to annual scales compared with gross methane production (Fig. S8), neglecting uptake introduces an uncertainty into the net carbon balance. In systems with extensive oxic zones, methane oxidation may offset a nontrivial fraction of emissions. Future work should explicitly represent methane uptake processes, particularly as marsh hydrology evolves under SLR and oxic–anoxic boundaries shift.

In terms of methane emission estimations, because our estimates are constrained by local observations, the study domain likely omits extremely high-emission habitats that elevate global averages. Future model refinements could incorporate additional field observations from diverse coastal tidal wetlands to improve empirical parameterization and strengthen data-driven training. Regionally, these coastal tidal marsh wetland methane emissions represent ~0.011–0.030% of the global methane flux from inland freshwater wetlands, while accounting for 0.072–0.12% of their total area (Zhang et al., 2021; Zhang et al., 2025). The relatively small contribution from coastal systems likely reflects the suppressing influence of saltwater intrusion on methane production.

Finally, our projections assume that marsh habitats keep pace with SLR, and that vegetation composition remains unchanged. This simplification overlooks critical feedback between marsh morphology, plant community shifts, and methane dynamics (Kirwan and

Megonigal, 2013; Cai et al., 2023). For example, vegetation changes can alter both belowground root oxygenation and organic matter supply, thereby modifying methane fluxes in ways not captured by our static framework. Incorporating dynamic marsh evolution models that account for sediment accretion, vegetation succession, and potential marsh drowning would provide a more realistic basis for long-term methane projections.

## Acknowledgments

XC is supported by an NSF OCE-PRF fellowship (grant no. 2403359). PR is supported by the U.S. Department of Energy (DE-SC0024709).

## Data availability

Scripts data analysis can be accessed at a https://doi.org/10.5281/zenodo.17916981 (Cai et al., 2025). Meta data used in this study can be found at https://doi.org/10.25573/serc.14227085. Remote sensing products is from https://gers.users.earthengine.app/view/tidalwetlandcover. Ocean model products are from NAAC (v1.0) built on the Semi-implicit Cross-scale Hydroscience Integrated System Model (SCHISM) (Cai et al., 2025).

## Author contributions

XC conceived the study and developed the overall concept. XC collected numerical data and performed data analysis. XY collected remote sensing data and performed data analysis. XC and PR developed the methane emission model. XC generated visualizations. PR provided guidance on application and discussion. XC prepared the initial manuscript draft. All authors contributed to manuscript writing and revisions.

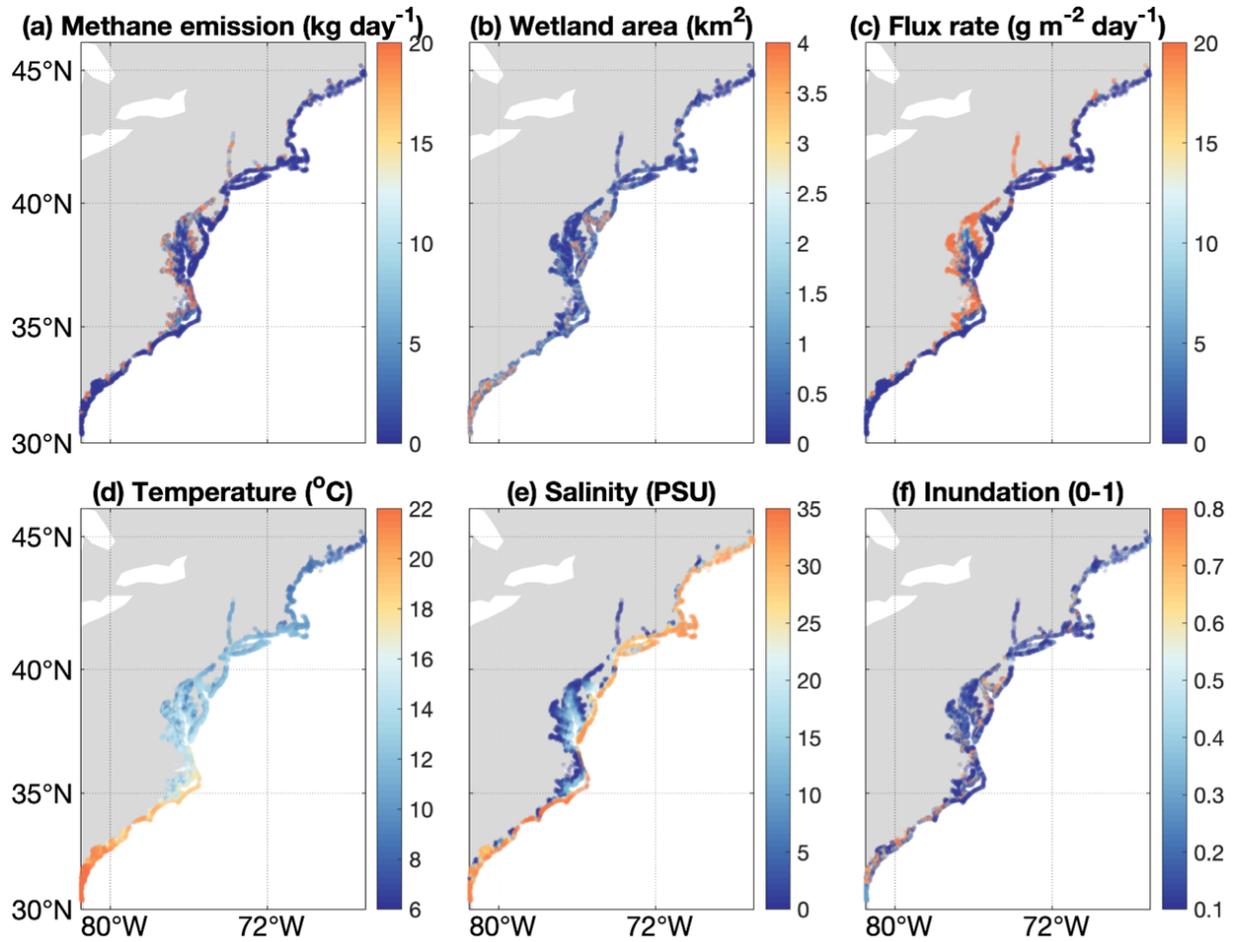

**Fig. S1**. Spatial distribution of 20-year mean (a) methane emissions per 9 km² sampling unit, (b) wetland area within each unit, (c) methane flux rates, (d) mean temperature, (e) mean salinity, and (f) inundation condition across coastal tidal marsh wetland regions.

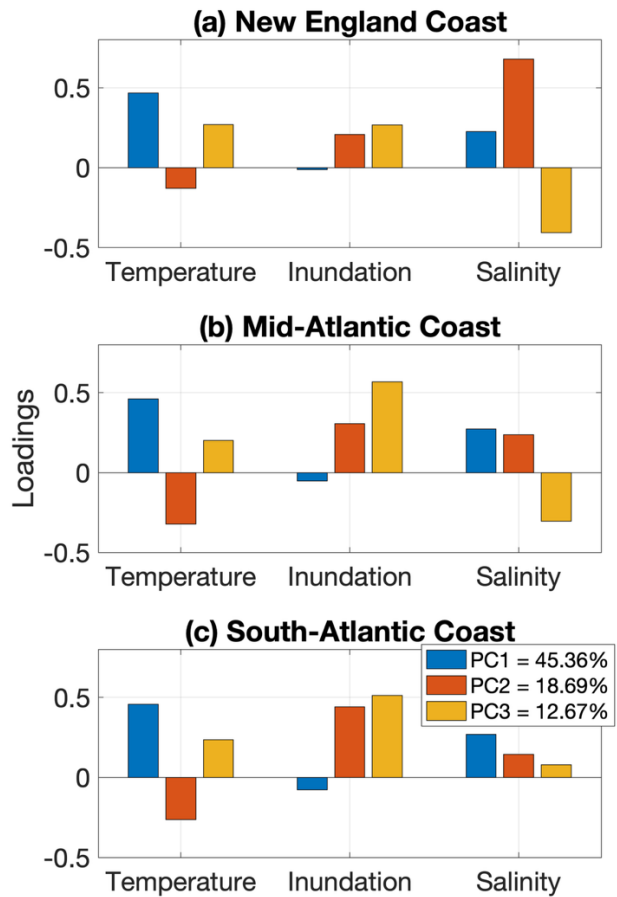

**Fig. S2**. Loadings of temperature, inundation, and salinity on the first three principal components (PCs) for (a) New England Coast, (b) Mid-Atlantic Coast, and (c) South Atlantic Coast. PC1 explains 45.36% of the variance, PC2 explains 18.69%, and PC3 explains 12.67% of the variance in the drivers.

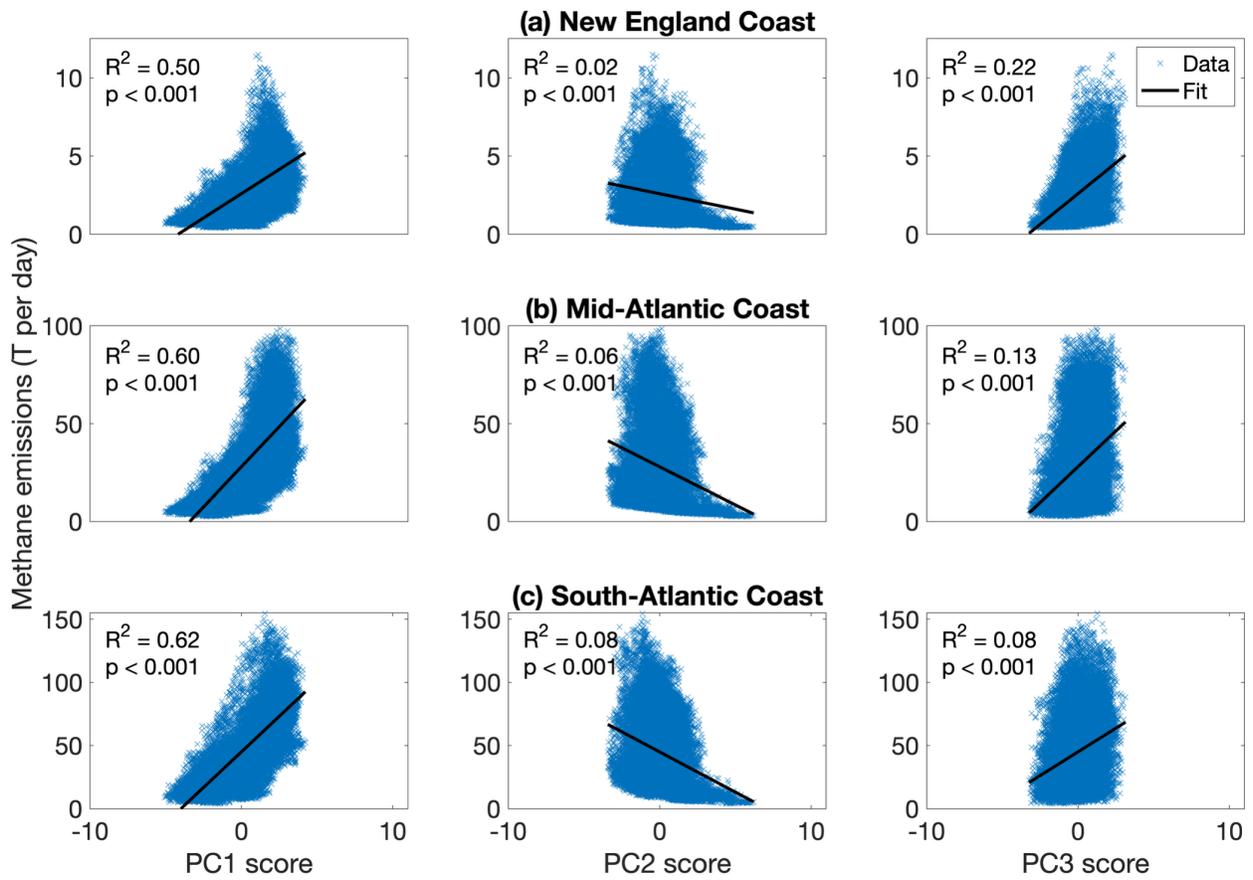

**Fig. S3**. Regression between estimated methane emissions to the first three principal components (PCs) for (a) New England Coast, (b) Mid-Atlantic Coast, and (c) South Atlantic Coast.

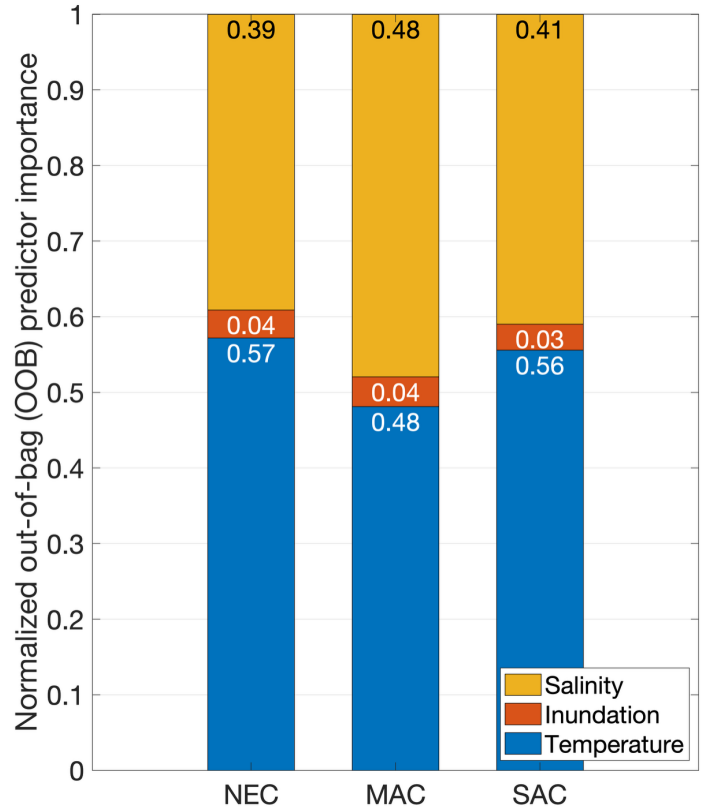

**Fig. S4**. Normalized out-of-bag (OOB) predictor importance of temperature, inundation, and salinity in estimating methane emissions across three coastal regions: the New England Coast (NEC), Mid-Atlantic Coast (MAC), and South Atlantic Coast (SAC). Predictor importance was derived from Random Forest regression models trained separately at each region. Values represent the relative contribution of each predictor to the overall model performance, normalized such that the total importance at each region sums to 1.

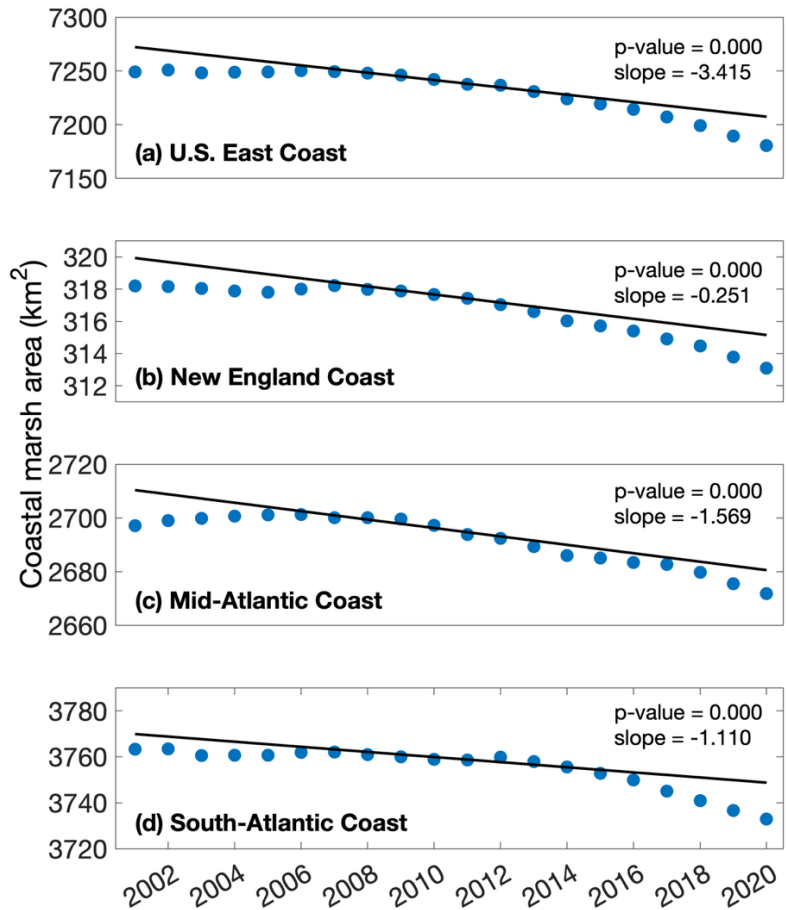

**Fig. S5**. Trends in tidal marsh area from 2001 to 2020 for the full study area (a), New England Coast (b), Mid-Atlantic Coast (c), and South Atlantic Coast (d). Blue circles show annual mean marsh area estimated from Landsat imagery, and black lines indicate Mann–Kendall trend results. The unit of the estimated slope is km$^2$ year$^{-1}$.

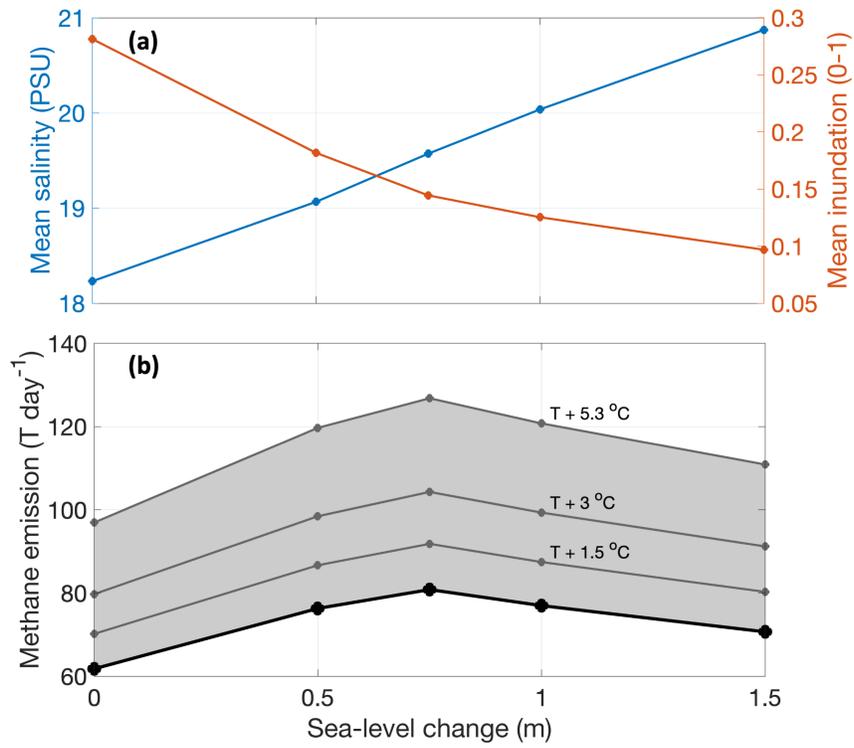

**Fig. S6**. (a) Response of mean salinity and inundation condition over the entire region upon different SLR conditions. (b) Predicted methane emissions under combined SLR and scenarios of increased temperature.

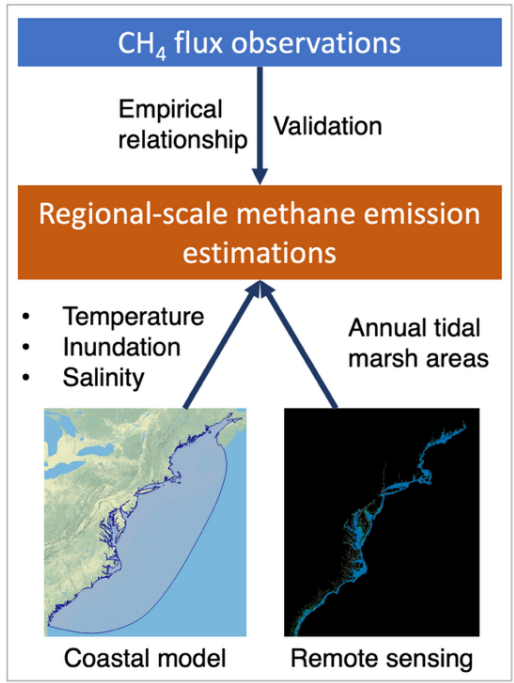

**Fig. S7.** Schematic of the model assembling.

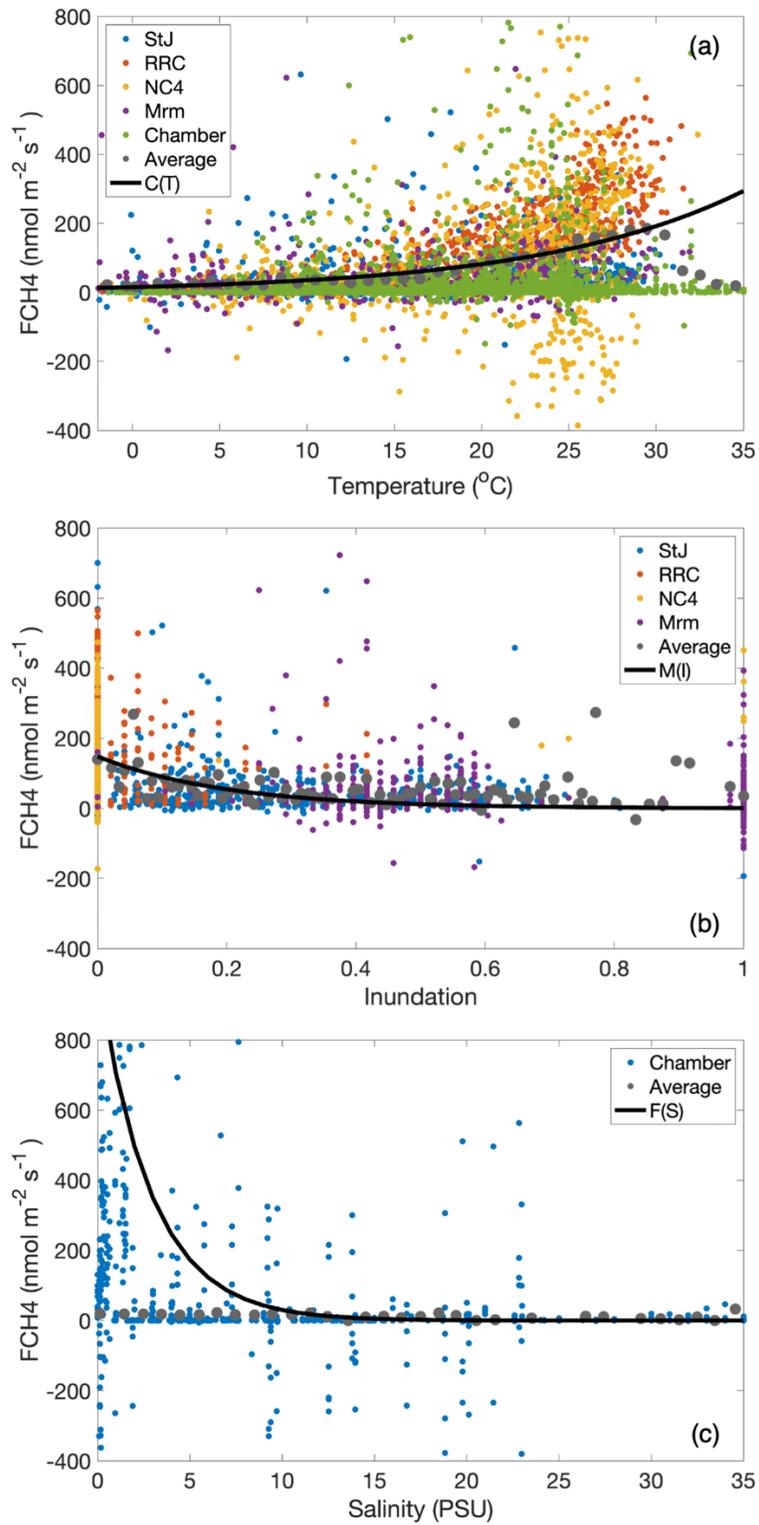

**Fig. S8**. Fitted exponential functions for C(T), M(I), and F(S) based on observational data from four eddy covariance towers and chamber measurements.

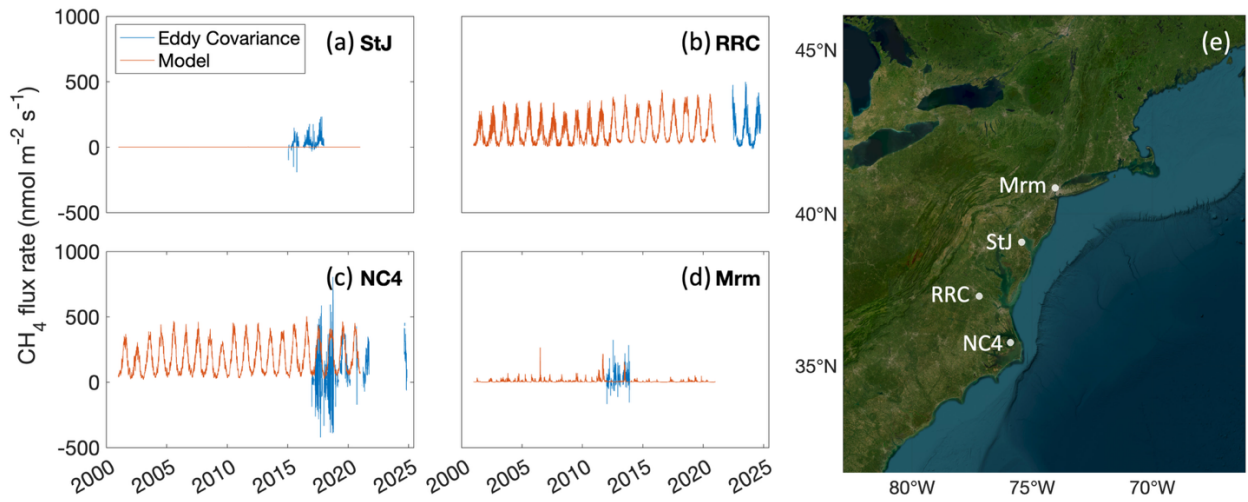

**Fig. S9**. (a-d) Comparison between modeled and observed methane emission rates at four eddy covariance stations. (e) Map showing the locations of the four stations.